\documentclass[twocolumn,prl,floatfix,citeautoscript,nofootinbib,superscriptaddress]{revtex4}
\usepackage{amsbsy}
\usepackage{latexsym,epsfig,graphicx}
\usepackage{dcolumn}
\usepackage{graphicx}
\usepackage{subfigure}
\usepackage{comment}
\usepackage{color}
\usepackage{bm}
\usepackage{mathrsfs}
\usepackage{amsfonts}
\usepackage{amsmath}
\usepackage{color}
\usepackage{amssymb}
\usepackage{xspace}
\usepackage{epstopdf}
\usepackage{tabularx}
\usepackage{longtable}
\usepackage[colorlinks=true, letterpaper=true, pdfstartview=FitV, linkcolor=blue, citecolor=blue, urlcolor=blue]{hyperref}
\usepackage[normalem]{ulem}
\usepackage{multirow}
\usepackage{esint}

\allowdisplaybreaks[4]
\pdfoutput=1

\begin{document}

\title{Non-Hermitian topological phase transitions for quantum spin Hall
insulators}
\author{Junpeng Hou}
\email{junpeng.hou@utdallas.edu}
\affiliation{Department of Physics, The University of Texas at Dallas, Richardson, Texas
75080-3021, USA}
\author{Ya-Jie Wu}
\affiliation{Department of Physics, The University of Texas at Dallas, Richardson, Texas
75080-3021, USA}
\affiliation{School of Science, Xi'an Technological University, Xi'an 710032, China}
\author{Chuanwei Zhang}
\email{chuanwei.zhang@utdallas.edu}
\affiliation{Department of Physics, The University of Texas at Dallas, Richardson, Texas
75080-3021, USA}

\begin{abstract}
The interplay between non-Hermiticity and topology opens an exciting avenue
for engineering novel topological matter with unprecedented properties.
While previous studies have mainly focused on one-dimensional systems or
Chern insulators, here we investigate topological phase transitions to/from
quantum spin Hall (QSH) insulators driven by non-Hermiticity. We show that a
trivial to QSH insulator phase transition can be induced by solely varying
non-Hermitian terms, and there exists \textit{exceptional edge arcs} in QSH
phases. We establish two topological invariants for characterizing the
non-Hermitian phase transitions: \textit{i}) with time-reversal symmetry,
the biorthogonal $\mathbb{Z}_{2}$ invariant based on non-Hermitian Wilson
loops, and \textit{ii)} without time-reversal symmetry, a biorthogonal spin
Chern number through biorthogonal decompositions of the Bloch bundle of the
occupied bands. These topological invariants can be applied to a wide class
of non-Hermitian topological phases beyond Chern classes, and provides a
powerful tool for exploring novel non-Hermitian topological matter and their
device applications.
\end{abstract}

\maketitle

{\color{blue}\emph{Introduction}}. Quantum spin Hall (QSH) insulator, a
topological phase of matter possessing quantized spin but vanishing charge
Hall conductance, has important applications in spintronics \cite%
{MurakamiDissipationless2003,AwschalomChallenges2007,BruneSpin2012} and was
widely studied in the past decade. It was pioneered by the celebrated
Kane-Mele model \cite{KaneQuantum2005} in graphene as a spinful enrichment
of the well-known Haldane model \cite{HaldaneModel1998} and later
generalized to other 2D materials (e.g., BHZ model \cite{BernevigQuantum2006}%
). The QSH insulator is topologically distinct from a trivial insulator by
its helical edge states, where different spins propagate along opposite
directions on the edge. In the presence of time-reversal (TR) symmetry, such
edge states correspond to a bulk topological invariant characterized by a $%
\mathbb{Z}_{2}$ index \cite{KaneZ22005}. Though being protected by TR
symmetry, the QSH effect survives under proper TR-broken term like exchange
field with the topological properties characterized by a $\mathbb{Z}$ spin
Chern number \cite{YangTime2011}.

The emergence of non-Hermitian physics provides an exciting platform for
engineering topological phases of matter with unprecedented properties that
are generally lacked in Hermitian systems. Many novel effects, such as
anomalous edge states, non-Bloch waves, biorthogonal bulk-edge
correspondence, etc. \cite%
{LeeTE2016,KunstFK2018,YaoS2018N,YaoS2018,LieuS2018,ShenH2018,KunstBiorthogonal2018,KawabataK2019,JinL2019,LinS2019,KawabataK2018}
have been revealed recently. On the experimental side, photonic lattices
\cite%
{MalzardS2015,StJeanP2017,BandresMA2018,PartoM2018,TakataK2018,ZhouH2018,CerjanA2018}%
, electronic circuits \cite{StehmannObservation2004,ChoiObservation2018},
and ultracold atoms \cite{Luo2019}, offer versatile platforms for realizing
non-Hermitian topological phases due to their high tunability and
controllability.

Previous studies on non-Hermitian topological matter have mainly focused on
two-dimensional (2D) Chern insulators or 1D systems such as SSH model or
Kitaev chain \cite{KawabataK2018}. Recently, a Kane-Mele model with
non-Hermitian Rashba spin-orbit interaction \cite{EsakiK2011} and a BHZ
model with non-Hermitian coupling terms \cite{KawabataK2019} have been
investigated. However the non-Hermiticity in these works cannot drive any
topological phase transition and the corresponding $\mathbb{Z}_{2}$
invariant does not depend on the non-Hermitian terms, leading to a plain $%
\mathbb{Z}_{2}$ index that is the same as that in Hermitian systems \cite%
{KawabataK2019}. Therefore a natural question is whether non-Hermiticity can
drive non-trivial topological phase transitions, e.g., from a trivial to a
non-Hermitian QSH insulator. If so, how can non-Hermiticity-driven
topological phase transitions be characterized? Does the $\mathbb{Z}_{2}$
index still apply and how do we define the bulk topological invariants?

In this Letter, we address these important questions by considering a
non-Hermitian generalization of Kane-Mele model with/without TR symmetry for
the realization of non-Hermiticity-driven QSH insulators. Our main results
are:

\textit{i)} We show that a topological phase transition from a trivial to a
QSH insulator with the emergence of purely real helical edge states can be
realized by solely tuning a TR-symmetric non-Hermitian term, which
originates from asymmetric Rashba spin-orbit interaction.

\textit{ii)} A transition from a QSH to trivial insulator can be driven by
another TR-symmetric non-Hermitian term, which splits the crossing of the
helical edges in the QSH phase into a pair of exceptional points that are
connected by \textit{exceptional edge arcs}.

\textit{iii)} In the presence of TR symmetry, we establish a biorthogonal $%
\mathbb{Z}_{2}$ index, which is defined by the parity of the winding of the
biorthogonal Wannier center derived from non-Hermitian Wilson loops. The
bulk biorthogonal $\mathbb{Z}_{2}$ invariant is consistent with the helical
edge states computed on a cylindrical geometry with zigzag boundary,
demonstrating the bulk-edge correspondence.

\textit{iv)} When the TR symmetry is broken, we establish the biorthogonal
spin Chern number, which is equivalent to the
biorthogonal $\mathbb{Z}_{2}$ invariant in the TR-symmetric region, to characterize non-Hermitian QSH insulators and their
phase transitions from/to a trivial or an integer quantum Hall insulator.

{\color{blue}\emph{Non-Hermitian QSH insulators with TR-symmetric}}. We
consider the Kane-Mele model on a 2D honeycomb lattice \cite%
{KaneQuantum2005,KaneZ22005}
\begin{eqnarray}
\mathcal{H}_{\text{KM}} &=&t\sum_{\langle i,j\rangle }c_{i}^{\dagger
}c_{j}+i\lambda _{\text{SO}}\sum_{\langle \langle i,j\rangle \rangle
}v_{ij}c_{i}^{\dagger }s_{z}c_{j}  \notag \\
&+&i\lambda _{\text{Rb}}\sum_{\langle i,j\rangle }c_{i}^{\dagger }(\bm{s}%
\times \bm{d}_{ij})_{z}c_{j}+\sum_{i}\lambda _{\nu }\xi _{i}c_{i}^{\dagger
}c_{i},  \label{KMMod}
\end{eqnarray}%
where $\bm{s}$ is Pauli matrix acting on the spin degree of freedom, and $t$
is the nearest neighbor hopping. $v_{ij}=\frac{2}{\sqrt{3}}(\bm{d}_{i}\times %
\bm{d}_{j})=\pm 1$ is defined through the unit vectors $\bm{d}_{i}$ and $%
\bm{d}_{j}$ along the transverse direction $\bm{d}_{ij}$ when particles hop
from site $j$ to $i$. $\xi _{i}=\pm 1$ applies on the sublattice degree of
freedom $\bm{\tau}$. The Bloch Hamiltonian in the momentum space can be
written as $H_{\text{KM}}\left( \bm{k}\right) =\sum_{a=1}^{5}d_{a}(\bm{k}%
)\Gamma ^{a}+\sum_{a<b}d_{ab}(\bm{k})\Gamma ^{ab}$, where the Dirac matrices
are defined as $\Gamma ^{a}=\big(\tau _{x}\otimes s_{0},\tau _{z}\otimes
s_{0},\tau _{y}\otimes s_{x},\tau _{y}\otimes s_{y},\tau _{y}\otimes s_{z}%
\big)$ with their commutators $\Gamma ^{ab}=\frac{1}{2i}[\Gamma ^{a},\Gamma
^{b}]$. $s_{0}$ and $\tau _{0}$ are identity matrices. The non-zero
coefficients $d_{a}(\bm{k})$ and $d_{ab}(\bm{k})$ \cite{KaneZ22005} are
listed in the Supplementary Materials \cite{mySupp}.

For simplicity, we consider only $\bm{k}$-independent non-Hermitian terms $%
d_{a}=i\lambda _{a}$ or $d_{ab}=i\lambda _{ab}$, where $\lambda _{a},\lambda
_{ab}\in \mathbb{R}$. The TR symmetry operator $\mathcal{T}=-i\tau
_{0}\otimes s_{y}K$ and $K$ is the complex conjugation, yielding $\mathcal{T}%
\Gamma ^{a}\mathcal{T}^{-1}=\Gamma ^{a}$ and $\mathcal{T}\Gamma ^{ab}%
\mathcal{T}^{-1}=-\Gamma ^{ab}$. Therefore a non-Hermitian term $i\lambda
_{a}\Gamma ^{a}$ ($i\lambda _{ab}\Gamma ^{ab}$) breaks (preserves) the TR
symmetry.

\begin{figure}[t]
\centering
\includegraphics[width=0.48\textwidth]{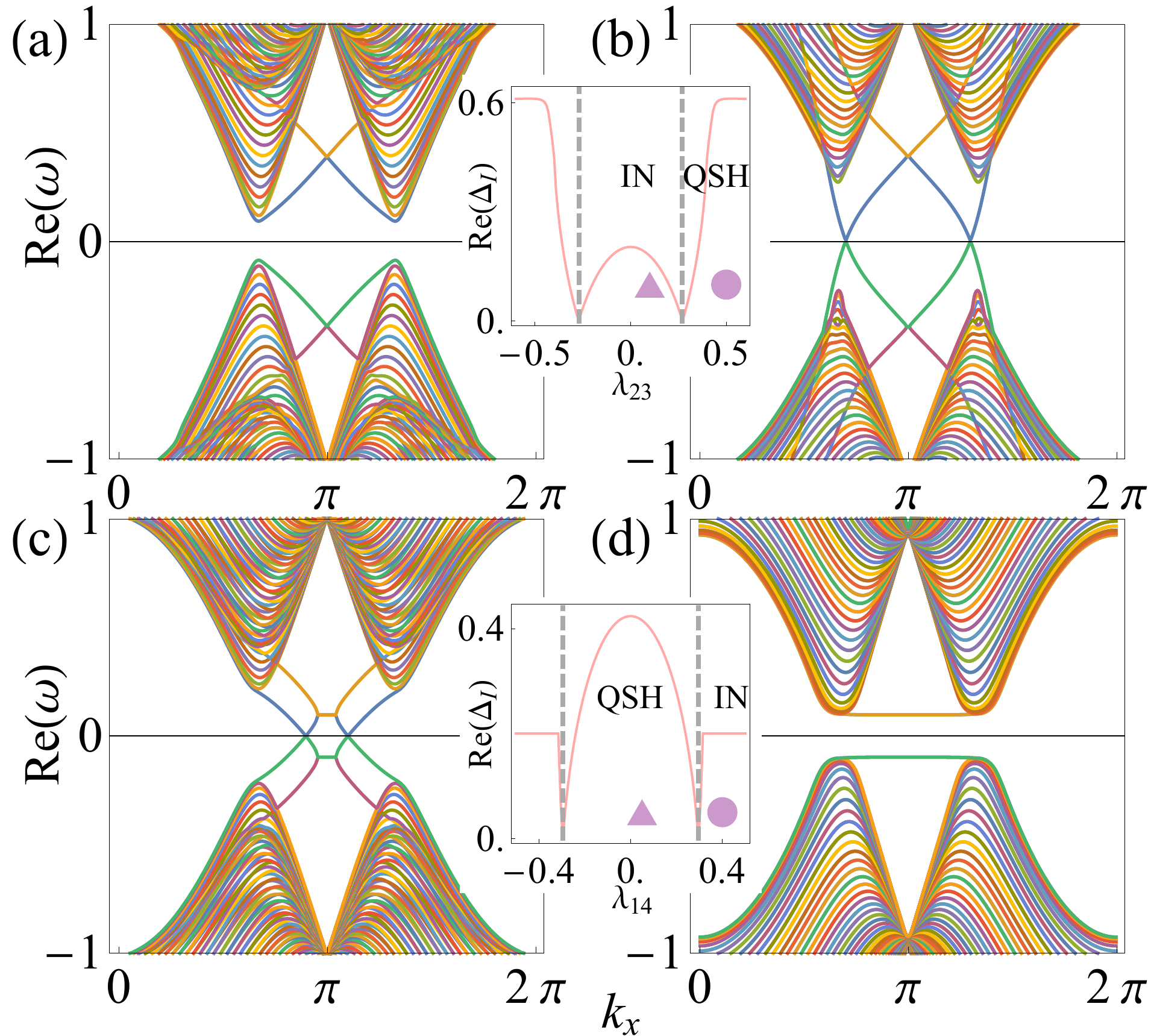}
\caption{Topological phase transitions driven by non-Hermiticity. (a,b) The
non-Hermitian term $i\protect\lambda _{23}\Gamma _{23}$ drives the system
from (a) trivial insulator ($\protect\lambda _{23}=0.1$) to (b) QSH
insulator ($\protect\lambda _{23}=0.5$). The panels show open-boundary
spectra on a zigzag ribbon. The parameters are $t=1$, $\protect\lambda _{%
\text{SO}}=0.06$, $\protect\lambda _{\protect\nu }=0.4$ and $\protect\lambda %
_{\text{Rb}}=0.05$. (c,d) The non-Hermitian term $i\protect\lambda %
_{14}\Gamma _{14}$ drives the system from (c) QSH insulator ($\protect%
\lambda _{14}=0.05$) to (d) trivial insulator ($\protect\lambda _{14}=0.4$)
with exceptional edge arcs. The parameters are the same as panels (a) and
(b) except $\protect\lambda _{\protect\nu }=0.1$ and $\protect\lambda _{%
\text{Rb}}=0$. The insets show the change of the real insulating gap Re($%
\Delta _{I}$) with respect to the non-Hermitian parameters. The dashed gray
lines represent gap closing points, \textquotedblleft IN\textquotedblright\
denotes the trivial insulator phase and the light purple triangles (disks)
denote the non-Hermitian parameters for panels on the left (right).}
\label{FigEdge}
\end{figure}

We first consider the TR-symmetric non-Hermitian Kane-Mele model $H_{\text{%
nH-KM}}=H_{\text{KM}}\left( \bm{k}\right) +i\lambda _{23}\Gamma ^{23}$. The
term $i\lambda _{23}\Gamma ^{23}$ mixes spins with non-Hermitian
nearest-neighbor hopping, yielding asymmetry in the Rashba spin-orbit
interaction along the bonds perpendicular to the zigzag edge. We start with
a trivial insulator with strong sublattice potential $\lambda _{\nu }=0.4$.
For a small $\lambda _{23}=0.1$, the open-boundary spectrum on a cylindrical
geometry with zigzag edge is plotted in Fig.~\ref{FigEdge}(a) and the edge
states do not cross the band gap, showing a trivial insulator. For weak $%
\lambda _{23}$, the insulating gap scales as $\Delta _{I}=6\sqrt{3}\lambda _{%
\text{SO}}-\sqrt{-\lambda _{23}^{2}+\lambda _{\nu }^{2}}-\sqrt{-\lambda
_{23}^{2}+\lambda _{\nu }^{2}+9\lambda _{\text{Rb}}^{2}}$ (the inset) with
the gap closing at $\lambda _{23}^{c}=\pm \sqrt{(4\lambda _{\text{SO}%
}\lambda _{\nu })^{2}-3(\lambda _{\text{Rb}}-12\lambda _{\text{SO}})^{2}}%
/(4\lambda _{\text{SO}})\approx \pm 0.27$ for the given parameters. With
increasing $\left\vert \lambda _{23}\right\vert >\left\vert \lambda
_{23}^{c}\right\vert $, the gap reopens and the system enters the QSH phase
with the emergence of helical edge states in the open-boundary spectrum at $%
\lambda _{23}=0.5$ (Fig.~\ref{FigEdge}(b)). Surprisingly, the edge states in
both regimes (trivial or topological) are purely real while the bulk
spectrums are complex \cite{mySupp}.

\begin{figure}[t]
\centering
\includegraphics[width=0.48\textwidth]{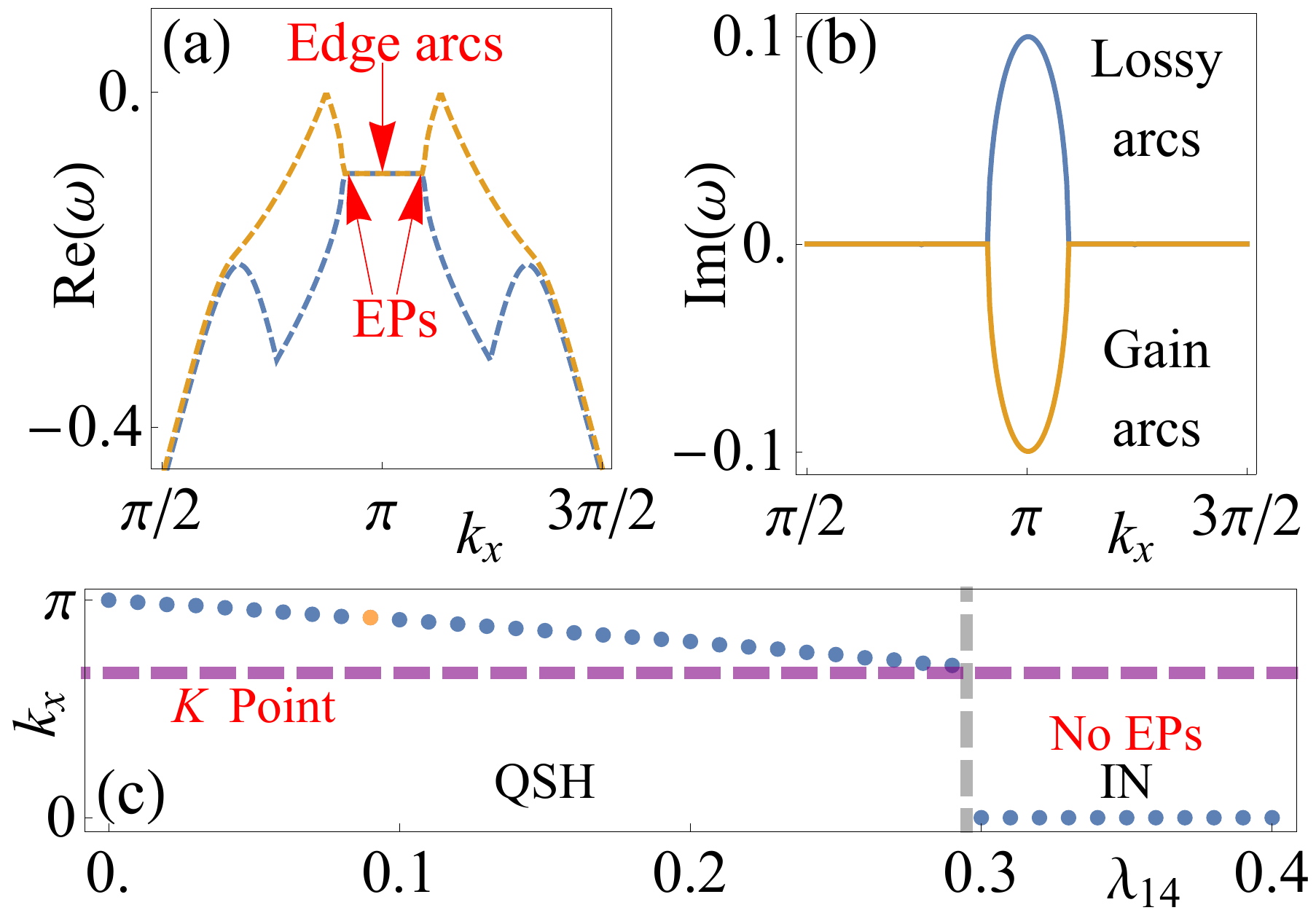}
\caption{Exceptional points and exceptional edge arcs. (a,b) Real and
imaginary parts of the edge states below the Fermi level. The parameters are
the same as Figs.~\protect\ref{FigEdge}(c,d) except $\protect\lambda %
_{14}=0.09$. The red arrows highlight the exceptional points (EPs) and edge
arcs. (c) Evolution of (right) exceptional point along $k_{x}$ with respect
to $\protect\lambda _{14}$.}
\label{FigEP}
\end{figure}

A different non-Hermitian term can also drive a phase transition from a QSH
to a trivial insulator with exceptional properties of the helical edge
states. We consider a QSH phase with $\lambda _{\nu }=0.1$ with vanishing
Rashba spin-orbit interaction and add the non-Hermitian term $i\lambda
_{14}\Gamma ^{14}$. This term splits the degeneracies of the helical edge
states at $k_{x}=\pi $ into two exceptional points (Fig.~\ref{FigEdge}(c)),
which are connected by two degenerate exceptional edge arcs with same real
but opposite imaginary parts. This is illustrated in Figs.~\ref{FigEP}(a,b),
where only the edge states below the Fermi level are plotted. The
exceptional points are developed between two components in the helical edge
state, which have opposite spins and chiralities (\textit{i.e.}, a
TR-symmetric pair). Along the exceptional edge arc, the spins are no longer
polarized.

The left and right exceptional points are symmetric to the $M$ point due to
TR symmetry. In Fig.~\ref{FigEP}(c), we plot the position of the right
exceptional point with respect to $\lambda _{14}$. It starts from the $M$
point ($k_{x}=\pi $) and moves almost linearly to $K$ point ($k_{x}=2\pi /3$%
), at which the insulating gap closes \cite{mySupp}. The eigenenergies at $K$%
/$K^{\prime }$ points are $\pm \lambda _{\nu }\pm \sqrt{27\lambda _{\text{SO}%
}^{2}-\lambda _{14}^{2}}$, therefore the gap closes at $\lambda
_{14}^{c}=\pm \sqrt{27\lambda _{\text{SO}}^{2}-\lambda _{\nu }^{2}}$. When
the band gap reopens for $\lambda _{14}>\lambda _{14}^{c}$, the system
becomes a trivial insulator, where the real insulating gap remains a
constant Re$(\Delta _{I})=2\lambda _{\nu }$ (see inset of Figs.~\ref{FigEdge}%
(c) and (d)). The exceptional edge arcs survive in the trivial insulator
with constant real energies while exceptional points vanish. Such
exceptional behaviors can be understood through a low-energy effective
Hamiltonian of the helical edge states \cite{mySupp}. We note that while
similar edge arcs in Fig.~\ref{FigEdge}(c) were observed previously \cite{KawabataK2019}, the
non-Hermiticity-driven topological phase transition was not investigated.

{\color{blue}\emph{Bulk biorthogonal $\mathbb{Z}_{2}$ invariant}}. The
topological phase transition and the emergence of helical edge states
indicate the change of the bulk topological invariant. In the Hermitian QSH
phase with TR symmetry, the bulk topology is characterized by a $\mathbb{Z}%
_{2}$ index \cite{KaneZ22005}, which is obtained from the phase winding $%
\mathcal{W}=\frac{1}{2\pi i}\oint_{\mathcal{L}}d\bm{k}\cdot \nabla _{\bm{k}%
}\log \mathcal{P}(\bm{k})$ along a closed path $\mathcal{L}$ that encircles
half of the Brillouin zone so that $\pm \bm{k}$ are not simultaneously
included. Here $\mathcal{P}(\bm{k})$ is the Pfaffian $\mathcal{P}(\bm{k})=%
\text{Pf}(\langle u^{m}(\bm{k})|\mathcal{T}|u^{n}(\bm{k})\rangle )$, where $%
m,n$ enumerates the occupied bands and $|u^{n}(\bm{k})\rangle $ is the
eigenvector. Because of the gauge-dependence of the Pfaffian, it is
difficult to extend the Pfaffian $\mathbb{Z}_{2}$ to general non-Hermitian systems for describing
non-Hermiticity-induced QSH insulators.

Here we establish a biorthogonal $\mathbb{Z}_{2}$ invariant for
non-Hermitian TR-invariant QSH insulators by developing a non-Hermitian
extension of the Wilson loop method \cite{RestaR1997}, which is equivalent
to Kane-Mele Pfaffian definition in Hermitian systems \cite{YuR2011}. The
biorthogonal Wilson line element is defined as
\begin{equation}
\lbrack G_{\alpha \beta }(\bm{k})]^{mn}={}_{\alpha }\langle u^{m}(\bm{k}%
+\Delta \bm{k})|u^{n}(\bm{k})\rangle _{\beta },\alpha \neq \beta ,
\end{equation}%
where $\Delta \bm{k}=(\bm{k}_{f}-\bm{k}_{i})/N$ is a small fraction on a
line constrained by two end points $\bm{k}_{i,f}$. $\alpha ,\beta =L,R$
denote the left and right eigenvectors, which are defined as $H(\bm{k})|u(%
\bm{k})\rangle _{R}=\omega (\bm{k})|u(\bm{k})\rangle _{R}$ and $H^{\dagger }(%
\bm{k})|u(\bm{k})\rangle _{L}=\omega ^{\ast }(\bm{k})|u(\bm{k})\rangle _{L}$%
. A normalization condition ${}_{\alpha }\langle u^{m}(\bm{k})|u^{n}(\bm{k}%
)\rangle _{\beta }=\delta ^{mn}$, $\alpha \neq \beta $ is imposed to form a
biorthogonal system \cite{ShenH2018}.

\begin{figure}[t]
\centering
\includegraphics[width=0.48\textwidth]{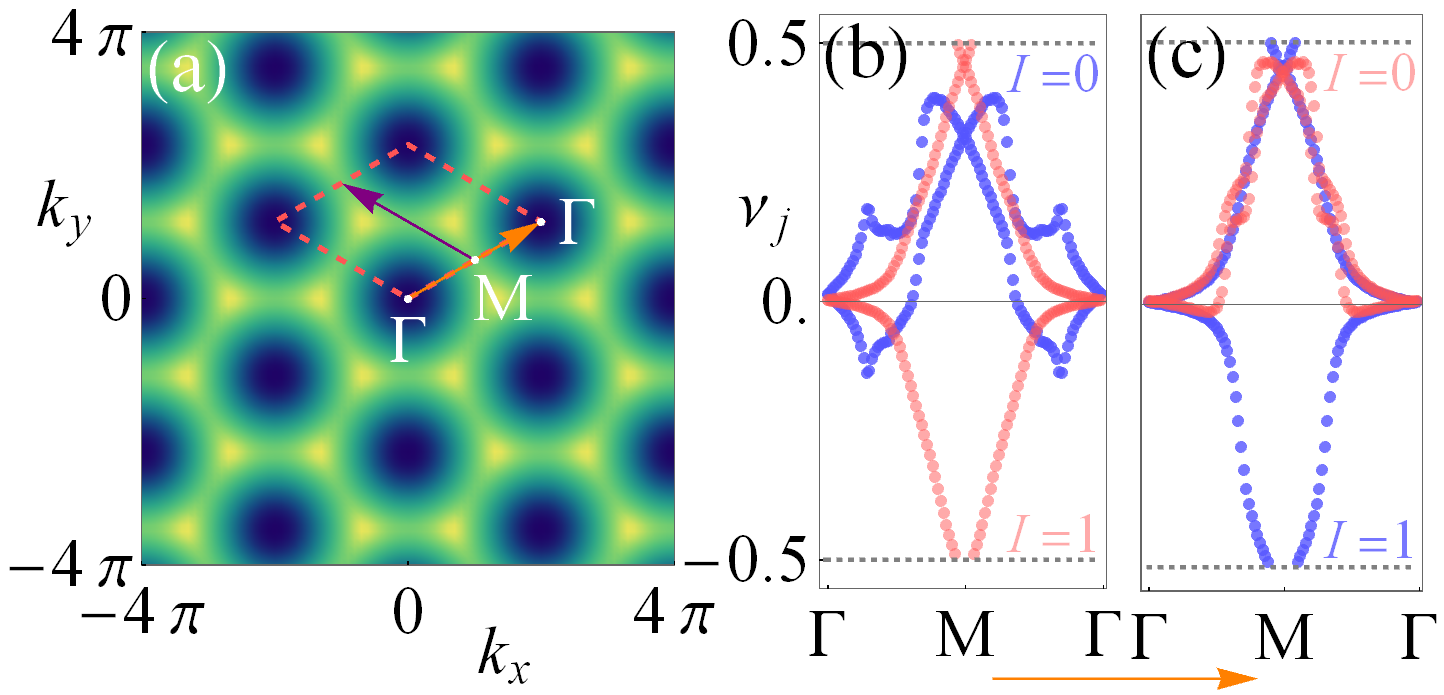}
\caption{Bulk biorthogonal $\mathbb{Z}_{2}$ invariant from non-Hermitian
Wilson loops. (a) The non-Hermitian Wilson loop in the Brillouin zone
(dashed parallelogram) is defined along the purple arrow while the base
point for each Wilson loop is given by the orange arrow. (b) and (c) The
biorthogonal Wannier centers plotted to varying base points, corresponding
to the cases in Figs.~\protect\ref{FigEdge}(a,b) and (c,d) respectively. The
blue (red) points represent weak (strong) non-Hermitian effects and the
biorthogonal $\mathbb{Z}_{2}$ index $I$ is labeled in each panel.}
\label{FigZ2}
\end{figure}

A path-ordered discrete Wilson line is defined as $\mathcal{W}_{\bm{k}%
_{i}\rightarrow \bm{k}_{f}}=G(\bm{k}_{f}-\Delta \bm{k})G(\bm{k}_{f}-2\Delta %
\bm{k})...G(\bm{k}_{i}+\Delta \bm{k})G(\bm{k}_{i})$ with%
\begin{equation}
\lbrack G(\bm{k})]^{mn}=\frac{1}{2}\left( [G_{LR}(\bm{k})]^{mn}+[G_{RL}(%
\bm{k})]^{mn}\right) .  \label{WLdef}
\end{equation}%
A Wilson loop $\mathcal{W}_{\bm{k}_{i}\rightarrow \bm{k}_{i}+\bm{T}}$,
\textit{i.e.}, a closed Wilson line, starts from the base point $\bm{k}_{i}$%
, and returns to $\bm{k}_{f}=\bm{k}_{i}+\bm{T}=\bm{k}_{i}$ after a period $\bm{T}$%
. The biorthogonal Wannier center $\nu _{j}(\bm{k}_{i})$ for each $\bm{k}%
_{i} $ is defined as the phase of the eigenvalues $E_{j}^{N}(\bm{k}%
_{i})=e^{i2\pi \nu _{j}(\bm{k}_{i})}$ of the Wilson loop through $\mathcal{W}%
_{\bm{k}_{i}\rightarrow \bm{k}_{i}+\bm{T}}(\bm{k})|v_{j}(\bm{k}_{i})\rangle
_{R}=E_{j}^{N}(\bm{k}_{i})|v_{j}(\bm{k}_{i})\rangle _{R}$ (in this work, $%
j=1,2$ for two lower occupied bands) and can be physically interpreted as
the relative position of the particle to the center of one unit cell \cite%
{mySupp}.

The biorthogonal $\mathbb{Z}_{2}$ invariant $I=\eta _{j}\mod 2$ is defined
by the winding $\eta _{j}=\frac{1}{2\pi }\oint_{\mathcal{C}}\nabla _{\bm{k}%
}\nu _{j}(\bm{k}_{i})\cdot d\mathbf{k}_{i}$ for each pair of biorthogonal
Wannier centers, where $\mathcal{C}$ is a loop for the base point $\bm{k}%
_{i} $ in the Brillouin zone. $I=1$ corresponds to the topological QSH
insulator with helical edge states for odd $\eta _{j}$, while $I=0$
corresponds to trivial insulator without any topologically protected edge
state for even $\eta _{j}$. When TR symmetry is preserved, the
winding of each biorthogonal Wannier center must come with its TR-symmetric
pair (either two 0 or a pair like $\pm 1$), therefore the biorthogonal $%
\mathbb{Z}_{2}$ invariant can always be defined.

To compute the biorthogonal Wannier center $\nu _{j}(\bm{k}_{i})$ on the
honeycomb lattice, we choose the Brillouin zone shown in Fig.~\ref{FigZ2}%
(a). The base point $\bm{k}_{i}$ is chosen along the orange arrow while the
purple arrow defines each non-Hermitian Wilson loop. The computed $\nu _{j}(%
\bm{k}_{i})$ for the QSH phase is displayed in Fig.~\ref{FigZ2}(b), where
the blue and red dots correspond to Figs.~\ref{FigEdge}(a,b)
respectively. There are two $\nu _{j}(\bm{k}_{i})$ for each color since
there are two occupied bands. The path $\mathcal{C}$ for the base point $%
\bm{k}_{i}$ starts from the $\Gamma $ point, sweeps through the $M$ point
and finally ends at another $\Gamma $ point, with $\nu _{j}(\bm{k}_{i})$
symmetric to $M$ point. At TR-symmetric points ($\Gamma $ and $M$), $\nu
_{j}(\bm{k}_{i})$ are degenerate as Kramers' pairs. Inversion symmetry
dictates that $\nu _{j}(\bm{k}_{i})$ must have opposite signs so that they
vanish at $\Gamma $. In the topological regime, $\nu _{j}(\bm{k}_{i})$
travel along different directions and show windings $\eta _{j}=\pm 1$,
yielding a biorthogonal $\mathbb{Z}_{2}$ index $I=1$ (red dots). In the
trivial insulator regime, $\nu _{j}(\bm{k}_{i})$ never crosses $\pm 1/2$ and
the winding vanishes, yielding $I=0$ (blue dots). Similarly,
the topological phase transition driven by $i\lambda _{14}\Gamma ^{14}$ in
Figs.~\ref{FigEdge}(c,d) is consistent with the change of bulk biorthogonal $%
\mathbb{Z}_{2}$ invariant from blue ($I=1$) to red ($I=0$) dots in Fig.~\ref%
{FigZ2}(c).

\begin{figure}[t]
\centering
\includegraphics[width=0.48\textwidth]{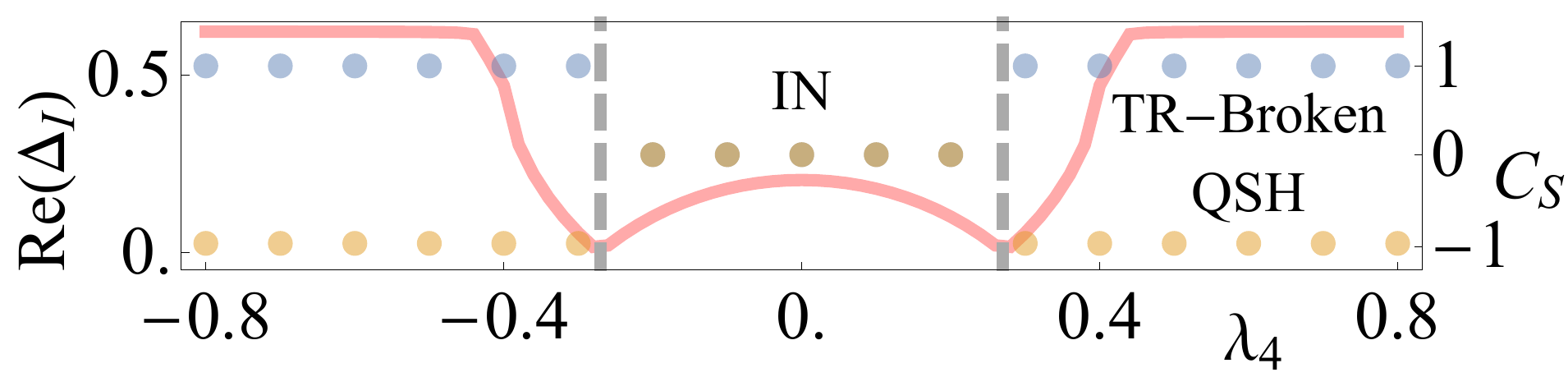}
\caption{Biorthogonal spin Chern number as the topological invariant for
TR-broken non-Hermitian QSH insulator. The parameters are the same as in
Figs.~\protect\ref{FigEdge}(a,b). The blue ($S=1/2$, spin \textquotedblleft
up") and orange ($S=-1/2$, spin \textquotedblleft down") dots are
biorthogonal spin Chern numbers $C_{S}$. The red curve represents the real
insulating gap.}
\label{FigSC}
\end{figure}

{\color{blue}\emph{Biorthogonal spin Chern number for TR-broken
non-Hermitian QSH insulators}}. When TR symmetry is broken, the windings of
Wannier centers may not come in pairs, therefore the biorthogonal $\mathbb{Z}%
_{2}$ invariant cannot be defined. Moreover, without TR symmetry, the system
reduces to the symmetry class classified by a $\mathbb{Z}$ topological
invariant \cite{KawabataK2018}. Here we consider a generalization of the
spin Chern number, which, in Hermitian systems, consists of a non-trivial
decomposition of a trivial Bloch bundle \cite{ProdanRobustness2009}.

We construct a biorthogonal $M$ matrix
\begin{equation}
\lbrack M(\bm{k})]^{mn}={}_{L}\langle u^{m}(\bm{k})|\tau _{0}\otimes
s_{z}|u^{n}(\bm{k})\rangle _{R},
\end{equation}%
whose diagonalization decomposes the mixed occupied bands into two spin
sectors (denoted by $S=\pm 1/2$) satisfying $M(\bm{k})|\psi _{S}(\bm{k}%
)\rangle _{R}=\omega _{S}|\psi _{S}(\bm{k})\rangle _{R}$. When the
eigenspectra $\omega _{S}$ of two spin sector are separable, we can define
the biorthogonal spin Chern number for each spin sector $C_{S,\alpha \beta }=%
\frac{1}{2\pi }\int d^{2}\bm{k}\cdot \mathcal{F}_{S,\alpha \beta }(\bm{k})$
through the Berry curvature $\mathcal{F}_{S,\alpha \beta }(\bm{k})=\nabla
\times \mathcal{A}_{S,\alpha \beta }(\bm{k})$, where the non-Abelian Berry
connection
\begin{equation}
\mathcal{A}_{S,\alpha \beta }(\bm{k})=-i{}_{\alpha }\langle \psi _{S}(\bm{k}%
)\cdot \bm{u}(\bm{k})|\partial _{\bm{k}}|\psi _{S}(\bm{k})\cdot \bm{u}(\bm{k}%
)\rangle _{\beta },
\end{equation}%
and $|\psi _{S}(\bm{k})\cdot \bm{u}(\bm{k})\rangle _{\beta }=\sum_{j}\psi
_{j}(\bm{k})|u^{j}(\bm{k})\rangle _{\beta }$. The summation of $j$ runs over
all occupied bands and $\psi _{j}(\bm{k})$ denotes the $j$-th component of
the eigenvector $|\psi _{S}(\bm{k})\rangle _{\beta }$. Previous studies have
shown that the Chern numbers defined through different Berry curvatures from
left or right eigenvectors are equivalent due to their gauge-invariant
nature \cite{ShenH2018}. Similar arguments apply here, and we denote $%
C_{S}=C_{S,\alpha \beta }$ hereafter.

In our context, a non-zero biorthogonal spin Chern number $C_{S}$ means
there is a chiral edge state of \textquotedblleft spin-$S$" with the
chirality determined by the sign of $C_{S}$. This is not generally true \cite%
{ProdanRobustness2009} but holds here because the underlying Haldane model
is a Chern insulator with the topological invariant quantized to 0 and $\pm 1
$ \cite{HaldaneModel1998}. In the trivial insulator phase $C_{\pm 1/2}=0$,
while $C_{\pm 1/2}=\pm 1$ in QSH phase. In the integer quantum Hall phase, $%
C_{\pm 1/2}=1$ or $-1$ \cite{mySupp}. Because $C_{S}$ is developed without
symmetry constraint, it can be applied in the TR-symmetric region. In fact,
the biorthogonal spin Chern number provides an equivalent description as the
biorthogonal $\mathbb{Z}_{2}$ invariant in the TR-symmetric region \cite%
{mySupp}.

The QSH phase survives even when TR-symmetry is broken except that there are
small backscatterings on the helical edge states, which open a small energy
gap on the edge states near the Fermi level \cite{YangTime2011}.
Such a TR-broken QSH phase can also be achieved directly from a trivial
phase through non-Hermiticity. For instance, we consider a term $i\lambda
_{4}\Gamma ^{4}$ that breaks TR symmetry, and start from a trivial insulator
phase as in Fig.~\ref{FigEdge}(a). This term also renders asymmetric Rashba
spin-orbit interaction and the gap scales similarly as the inset in Figs.~%
\ref{FigEdge}(a,b). With increasing $|\lambda _{4}|$, the gap closes and
then reopens, leading to a TR-broken QSH phase, as shown in Fig.~\ref{FigSC}. The biorthogonal
spin Chern numbers $C_{S}$ for both trivial insulator and TR-broken QSH
phases are computed, which are consistent with the open-boundary spectra
(see Fig.~\ref{FigEdgeTRbk} in Supplemental Materials \cite{mySupp}). In
Hermitian systems, a topological phase transition from a TR-broken QSH phase
to an interger quantum Hall phase can be driven by a real exchange field $%
\lambda _{\mu }\Gamma ^{34}$. Such a phase transition still exists in the
non-Hermitian region and can be characterized by the biorthogonal spin Chern
number \cite{mySupp}.

{\color{blue}\emph{Conclusion and Discussion}}. In summary, we have
demonstrated that QSH insulators and their phase transitions from/to trivial
insulators can be driven by non-Hermiticity and showcased the exceptional edge arcs under topological phase transition. While our discussion focuses on non-Hermitian
Kane-Mele model, the developed topological invariants, \textit{i.e.}, the
biorthogonal $\mathbb{Z}_{2}$ index and spin Chern number $C_{S}$, are
applicable to other QSH models like non-Hermitian BHZ model. The
biorthogonal $\mathbb{Z}_{2}$ index may be further generalized to
characterize 3D non-Hermitian topological insulators, which needs further
investigation. The biorthogonal $\mathbb{Z}_{2}$ (and $\mathbb{Z}$)
topological invariants provide a powerful tool for characterizing wide
classes of non-Hermitian topological matters and pave the way for exploring
their device applications.

\begin{acknowledgments}
\textbf{Acknowledgments:} This work was supported by Air Force Office of
Scientific Research (FA9550-16-1-0387), National Science Foundation
(PHY-1806227), and Army Research Office (W911NF-17-1-0128). Y.W. was also
supported in part by NSFC under the grant No. 11504285 and the Scientific
Research Program Funded by Natural Science Basic Research Plan in Shaanxi
Province of China (Program No. 2018JQ1058).
\end{acknowledgments}

\appendix

\newpage \clearpage
\onecolumngrid
\appendix

\section{Supplemental Materials for ``Non-Hermitian topological phase transitions for quantum spin Hall
insulators''}

\setcounter{table}{0} \renewcommand{\thetable}{S\arabic{table}} %
\setcounter{figure}{0} \renewcommand{\thefigure}{S\arabic{figure}}

The Supplemental Materials provide more details on imaginary bands,
exceptional points on helical edge states, non-Hermitian Wilson line in
thermodynamic limits, biorthogonal spin Chern number as topological
invariant for TR-symmetric non-Hermitian QSH insulators, edge states of
non-Hermiticity-driven TR-broken QSH phase and non-Hermitian IQH phase.

\subsection{Imaginary bands of open-boundary spectra}

For the convenience of the reader, we list the non-zero coefficients of the
Kane-Mele model \cite{KaneZ22005}
\begin{eqnarray}
d_{1} &=&t\left( 1+2\cos \left( \frac{1}{2}k_{x}\right) \cos \left( \frac{%
\sqrt{3}}{2}k_{x}\right) \right) ,  \notag \\
d_{2} &=&\lambda _{\nu },  \notag \\
d_{3} &=&\lambda _{\text{Rb}}\left( 1-\cos \left( \frac{1}{2}k_{x}\right)
\cos \left( \frac{\sqrt{3}}{2}k_{x}\right) \right) ,  \notag \\
d_{4} &=&-\sqrt{3}\lambda _{\text{Rb}}\ \sin \left( \frac{1}{2}k_{x}\right)
\sin \left( \frac{\sqrt{3}}{2}k_{x}\right) ,  \notag \\
d_{12} &=&-2t\cos \left( \frac{1}{2}k_{x}\right) \sin \left( \frac{\sqrt{3}}{%
2}k_{x}\right) ,  \notag \\
d_{15} &=&2\lambda _{\text{SO}}\ \left( \sin k_{x}-2\cos \left( \frac{1}{2}%
k_{x}\right) \cos \left( \frac{\sqrt{3}}{2}k_{x}\right) \right) ,  \notag \\
d_{23} &=&-\lambda _{\text{Rb}}\ \cos \left( \frac{1}{2}k_{x}\right) \sin
\left( \frac{\sqrt{3}}{2}k_{x}\right) ,  \notag \\
d_{24} &=&\sqrt{3}\lambda _{\text{Rb}}\ \sin \left( \frac{1}{2}k_{x}\right)
\cos \left( \frac{\sqrt{3}}{2}k_{x}\right) ,  \notag \\
d_{34} &=&\lambda _{\mu },  \notag
\end{eqnarray}%
where the last term represents a TR-broken exchange field and is introduced
in TR-broken QSH insulators \cite{YangTime2011}.

\begin{figure}[h]
\centering
\includegraphics[width=0.48\textwidth]{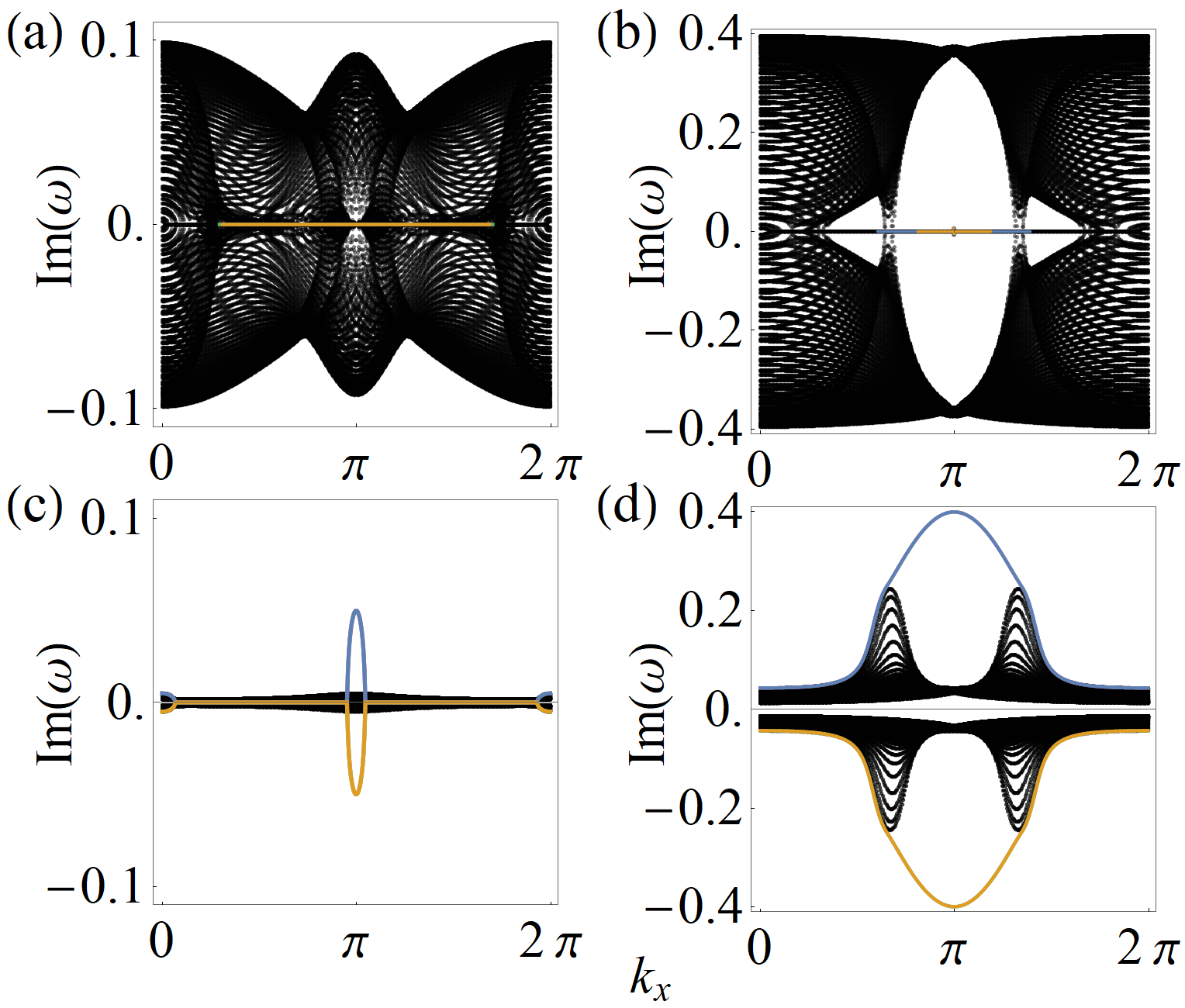}
\caption{Imaginary bands in TR-symmetric non-Hermitian Kane-Mele model.
(a-d) Imaginary parts of the spectra shown in Figs.~\protect\ref{FigEdge}%
(a-d) respectively. The edge states are plotted with the same color as those
in Fig.~\protect\ref{FigEdge} and we only plot the purely real edge states
in panels (a) and (b).}
\label{FigEdgeSupp}
\end{figure}

In the main text, we examine two topological phase transitions in the
TR-symmetric non-Hermitian Kane-Mele model. Only real parts of the
open-boundary spectra are plotted in the main text (see Fig.~\ref{FigEdge}).
The imaginary bands are plotted here in Fig.~\ref{FigEdgeSupp}.

For the topological phase transition from trivial to QSH insulators driven
by non-Hermitian term $i\lambda _{23}\Gamma ^{23}$, the edge states in both
phases are (mostly) purely real while the bulk spectrum is complex, as shown
in Figs.~\ref{FigEdgeSupp}(a,b). The helical edge states are separated
from the bulk bands in the entire complex plane.

Different from asymmetric Rashba spin-orbit interaction $i\lambda
_{23}\Gamma ^{23}$, $i\lambda _{14}\Gamma ^{14}$ term represents a
non-Hermitian next-nearest-neighbor hopping that only mixes spins. It splits
the edge crossing into a pair of exceptional points, which are connected by
an exceptional edge arc with same real energy and opposite imaginary
energies, as shown in Fig.~\ref{FigEdgeSupp}(c). Outside the exceptional
edge arc, the edge state spectrum is purely real. In the trivial insulator
phase, the entire spectrum becomes complex (see Fig.~\ref{FigEdgeSupp}(d)).

\subsection{Bulk bands and low-energy theory of exceptional edge arcs}

A non-zero $\Gamma _{14}$ term couples different spins while preserves the
sublattice degree of freedom. With a real coefficient, $\lambda _{14}\Gamma
_{14}$ breaks the TR symmetry and opens a finite gap on the helical edges.
There is no gap closing or topological phase transition. In contrast, the
imaginary term $i\lambda _{14}\Gamma _{14}$ generates a pair of exceptional
points on the gapless helical edge states that cross at TR-symmetric points.
Two similar terms are $i\lambda _{13}\Gamma _{13}$ and $i\lambda _{35}\Gamma
_{35}$, which can also induce topological phase transitions through
splitting the edge crossings into exceptional points.

\begin{figure}[h]
\centering
\includegraphics[width=0.6\textwidth]{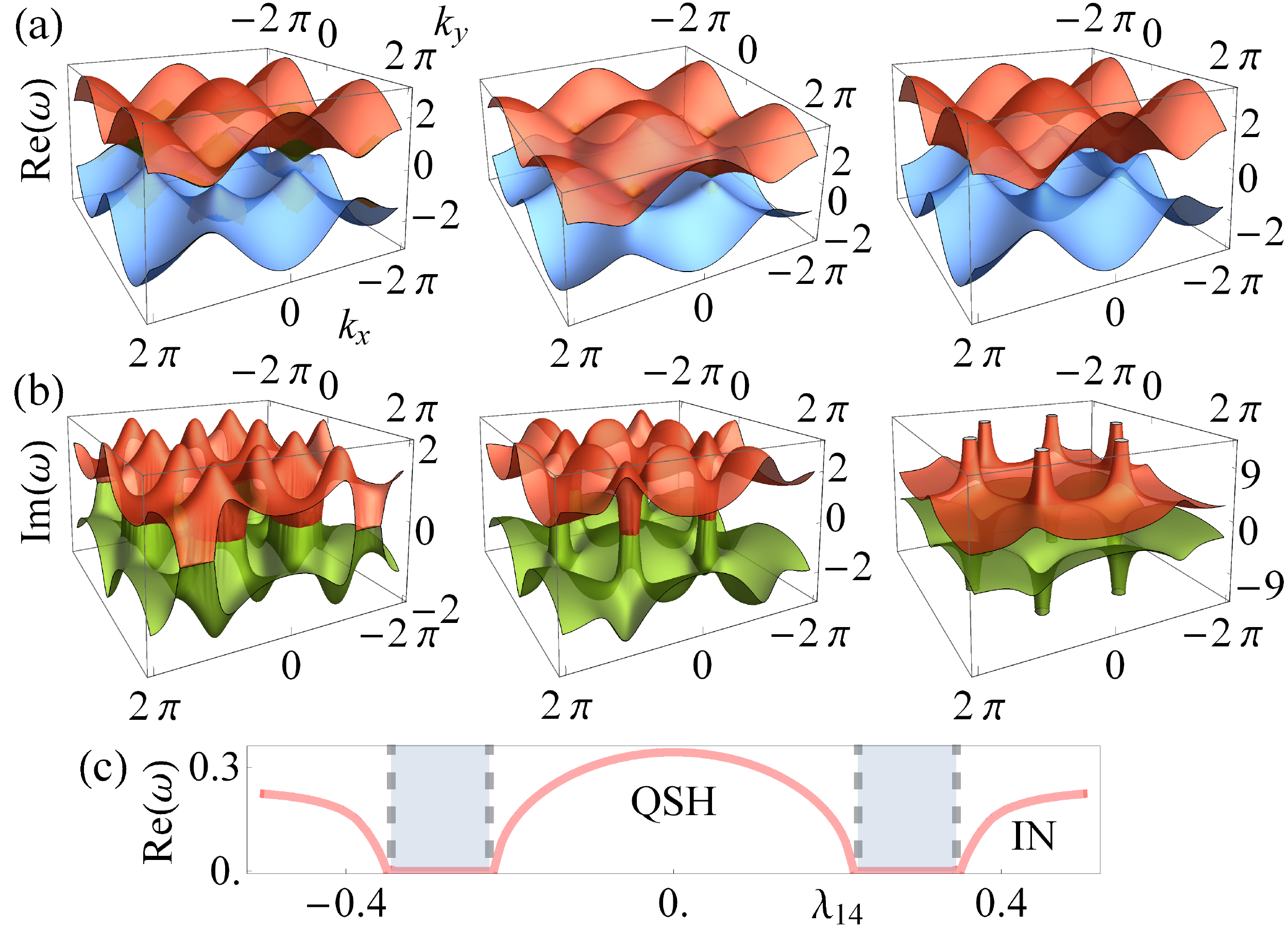}
\caption{(a,b) Real and imaginary bands in momentum space across the
phase transition. The parameters are $\protect\lambda _{14}=0.2$, $0.295$
(gap closing point) and $0.4$. (c) Topological phase transition driven by $i%
\protect\lambda _{14}\Gamma _{14}$ with nonzero Rashba spin-orbit
interaction. The parameters are the same as those in Fig.~\ref{FigEP} in the main text
except $\protect\lambda _{\text{Rb}}=0.05$. The light-blue-coded areas
correspond to the gapless phases.}
\label{FigEPSupp}
\end{figure}

In the main text, we have described the properties of exceptional edge arcs.
Here we plot the change of the bulk band spectrum across the phase
transition with the band gap closing in Figs.~\ref{FigEPSupp}(a) and (b).
The gap closing happens in the complex plane, meaning both real and
imaginary parts of the eigenvalues must vanish.

A similar picture holds when Rashba spin-orbit interaction exists or TR
symmetry is broken by a small exchange field $\lambda _{34}\Gamma ^{34}$.
The Rashba term turns the real edge states outside the exceptional edge arcs
into complex states and the exchange field simply shifts the exceptional
points to different directions. The topological phase transition driven by $%
i\lambda _{14}\Gamma ^{14}$ with small Rashba spin-orbit interaction (the
system is still topological when $\lambda _{14}=0$) is plotted in Fig.~\ref%
{FigEPSupp}(c). The critical points for the phase transition become gapless
phases and the biorthogonal $\mathbb{Z}_{2}$ index developed in the main
text still applies. Finally, we note that one can change the open-boundary
direction and observe the same physics along $k_{y}$.

To understand how a non-Hermitian term induces exceptional points on the
helical edge states, we consider a low-energy effective Hamiltonian%
\begin{equation}
H_{\text{edge}}=k_{x}\Gamma _{15}+\lambda _{\nu }\Gamma _{2}+i\lambda
_{14}\Gamma _{14},
\end{equation}%
which preserves the TR symmetry. The first term $k_{x}\Gamma _{15}$
describes a Dirac fermion with the four-fold degeneracy at $k_{x}=0$. The
second term $\lambda _{\nu }\Gamma _{2}$ lifts the degeneracy and renders 4
edge crossings, two of which locate at the Fermi level with opposite $k_{x}$
while the other two at $k_{x}=0$ with opposite energies. Such a band
structure resembles the edge spectrum in a QSH insulator. The last term $%
i\lambda _{14}\Gamma _{14}$ stretches each band crossing below and above the
Fermi level into two exceptional points. Diagonalize the above Hamiltonian,
we obtain
\begin{eqnarray}
&&\omega _{\pm ,-}=\pm \sqrt{k_{x}^{2}-\lambda _{14}^{2}}-\lambda _{\nu
},\omega _{\pm ,+}=\pm \sqrt{k_{x}^{2}-\lambda _{14}^{2}}+\lambda _{\nu }, \\
&&|\psi _{\pm ,-}\rangle _{R}=(0,-\frac{k_{x}-\omega _{\pm ,-}-\lambda _{\nu
}}{\lambda _{14}},0,1)^{T},~|\psi _{\pm ,+}\rangle _{R}=(-\frac{k_{x}+\omega
_{\pm ,+}-\lambda _{\nu }}{\lambda _{14}},0,1,0)^{T},
\end{eqnarray}%
where $\omega _{\pm ,\pm }$ denote four eigenvalues and $|\psi _{\pm ,\pm
}\rangle _{R}$ are the corresponding right eigenvectors. Both the
eigenvalues and the right eigenvectors collapse at the exceptional points $%
k_{x}=\pm \lambda _{14}$ (so do the left eigenvectors). In general,
exceptional points are not protected by TR symmetry. If the TR symmetry is
broken with an exchange field $\lambda _{34}\Gamma _{34}$, the exceptional
points merely shift their positions to $k_{x}=\pm \lambda _{14}-\lambda _{34}
$ (for the pair above the Fermi level) or $k_{x}=\pm \lambda _{14}+\lambda
_{34}$ (for the pair below the Fermi level). We also note that the
topological phase transition studied in Figs.~\ref{FigEdge}(c,d) still
occurs in TR-broken non-Hermitian QSH insulators.

\subsection{Biorthogonal Wilson line in thermodynamic limit}

In non-Hermitian systems, the Berry connection cannot be solely defined
through the right eigenvectors. A proper way to define a purely real Berry
connection involves both left and right eigenvectors
\begin{equation}
\mathcal{A}(\bm{k})=\frac{1}{2}\left( \mathcal{A}_{LR}(\bm{k})+\mathcal{A}%
_{RL}(\bm{k})\right) ,
\end{equation}%
where $[\mathcal{A}_{\alpha \beta }(\bm{k})]^{mn}=-i{}_{\alpha }\langle
u^{m}(\bm{k})|\partial _{\bm{k}}u^{n}(\bm{k})\rangle _{\beta }$, $\alpha
\neq \beta $ is the biorthogonal non-Abelian Berry connection. $\mathcal{A}(%
\bm{k})$ is real since $\mathcal{A}_{LR}(\bm{k})=\mathcal{A}_{RL}^{\ast }(%
\bm{k})$.

To justify our definition of the non-Hermitian Wilson line element%
\begin{equation}
\lbrack G(\bm{k})]^{mn}=\frac{1}{2}\left( [G_{LR}(\bm{k})]^{mn}+[G_{RL}(%
\bm{k})]^{mn}\right)
\end{equation}%
in the main text, we need show that the non-Hermitian Wilson line gives the
desired non-Hermitian Berry phase in the thermodynamic limits, similar to
the Hermitian cases \cite{RestaR1997}. We expand each element $[G_{\alpha
\beta }(\bm{k})]^{mn}={}_{\alpha }\langle u^{m}(\bm{k}+\Delta \bm{k})|u^{n}(%
\bm{k})\rangle _{\beta }$ to the first order (assuming $N$ is very large)
\begin{equation}
\lbrack G_{\alpha \beta }(\bm{k})]^{mn}={}_{\alpha }\langle u^{m}(\bm{k}%
)|u^{n}(\bm{k})\rangle _{\beta }+(\Delta \bm{k}){}_{\alpha }\langle \partial
_{\bm{k}}u^{m}(\bm{k})|u^{n}(\bm{k})\rangle _{\beta }.
\end{equation}%
Due to the normalization condition ${}_{\alpha }\langle u^{m}(\bm{k})|u^{n}(%
\bm{k})\rangle _{\beta }=\delta ^{mn}$ \cite{ShenH2018}, we have ${}_{\alpha
}\langle \partial _{\bm{k}}u^{m}(\bm{k})|u^{n}(\bm{k})\rangle _{\beta
}=-{}_{\alpha }\langle u^{m}(\bm{k})|\partial _{\bm{k}}u^{n}(\bm{k})\rangle
_{\beta }$. The biorthogonal Wilson line element can be rewritten as
\begin{equation}
\lbrack G_{\alpha \beta }(\bm{k})]^{mn}=\delta ^{mn}-i(\Delta \bm{k})[%
\mathcal{A}_{\alpha \beta }(\bm{k})]^{mn},
\end{equation}%
yielding
\begin{equation}
\lbrack G(\bm{k})]^{mn}=\delta ^{mn}-i(\Delta \bm{k})[\mathcal{A}(\bm{k}%
)]^{mn}.
\end{equation}

The non-Hermitian Wilson loop from $\bm{k}_{i}$ to $\bm{k}_{f}$ is defined
through a path-ordered multiplication
\begin{equation}
\mathcal{W}_{\bm{k}_{i}\rightarrow \bm{k}_{f}}=\prod_{j=1}^{N}\left(
I_{0}-i(\Delta \bm{k})\mathcal{A}(\bm{k}+j\Delta \bm{k})\right) ,
\end{equation}%
where $I_{0}$ is the identity matrix. Under the thermodynamic limit $%
N\rightarrow \infty $, it gives the exponential of the non-Hermitian Berry
phase
\begin{equation}
\lim_{N\rightarrow \infty }\mathcal{W}_{\bm{k}_{i}\rightarrow \bm{k}%
_{f}}=e^{-i\int_{\bm{k}_{i}}^{\bm{k}_{f}}\mathcal{A}(\bm{k})\cdot d\bm{k}}.
\end{equation}
This equation demands that the non-Hermitian Wilson loop must be unitary in
the thermodynamic limit so that the biorthogonal Wannier center can be
defined.

Since the Berry phase represents the electronic contribution to the
dielectric polarization in solid state, the biorthogonal Wannier center $\nu
_{j}(\bm{k})$ can be physically interpreted as the relative position of the
particle to the center of one unit cell with the polarization
\begin{equation}
p=-\frac{i}{2\pi }\log \det (\mathcal{W}_{\bm{k}_{i}\rightarrow \bm{k}_{i}+%
\bm{T}}).
\end{equation}

\subsection{Biorthogonal spin Chern number in TR-symmetric non-Hermitian QSH
insulators}

In the main text, we claim that biorthogonal spin Chern number also works
when TR-symmetry is preserved. Here we use the biorthogonal spin Chern
number to characterize two topological phase transitions studied in Fig.~\ref%
{FigEdge}.

\begin{figure}[h]
\centering
\includegraphics[width=0.48\textwidth]{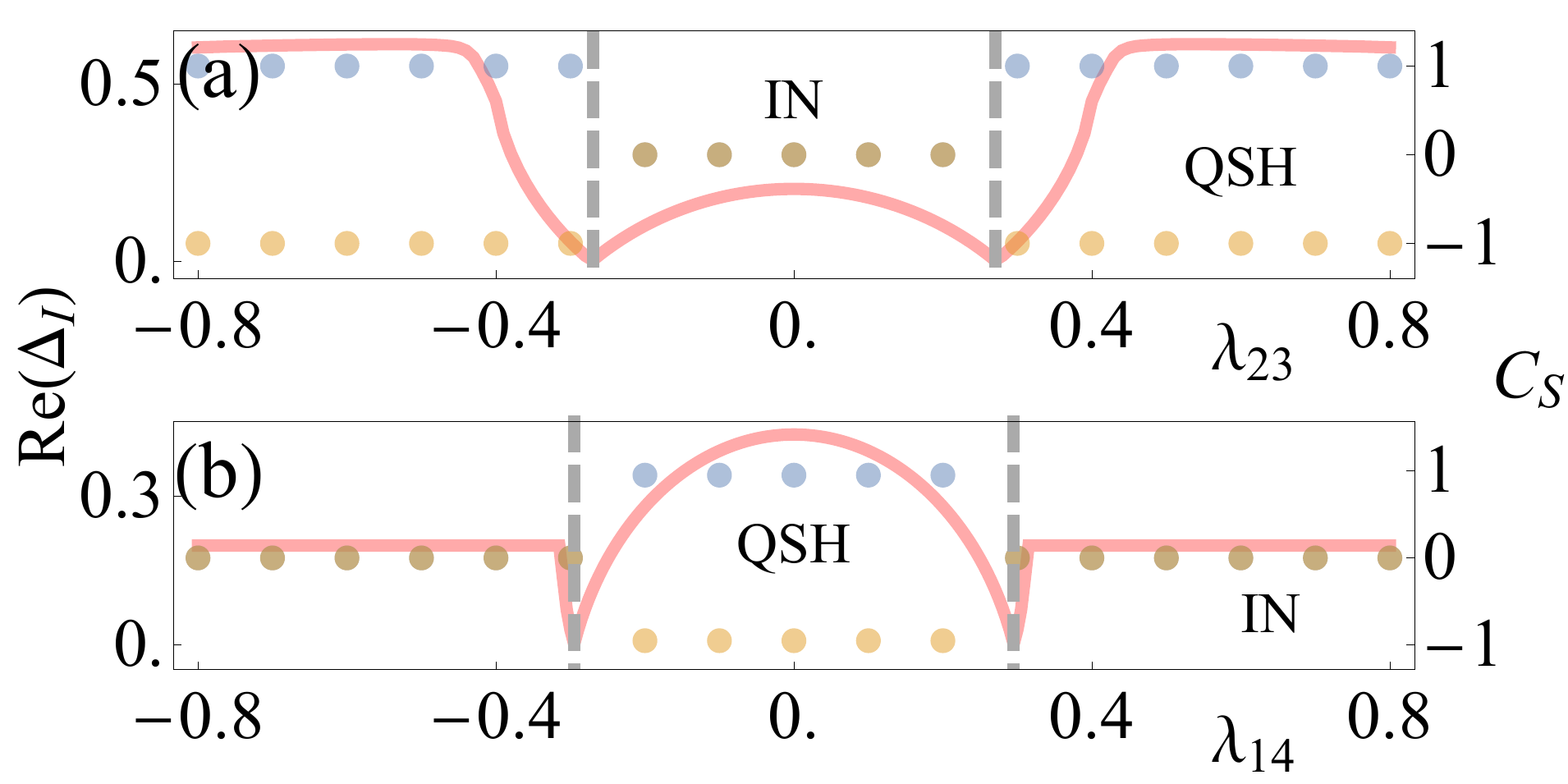}
\caption{Biorthogonal spin Chern number as topological invariant for
TR-symmetric non-Hermitian QSH insulators. (a,b) Biorthogonal spin
Chern numbers computed for the TR-symmetric models in Figs.~\protect\ref%
{FigEdge}(a,b) and (c,d), respectively.}
\label{FigSCTR}
\end{figure}

The first topological phase transition from a trivial insulator to a QSH
insulator is driven by the non-Hermitian term $i\lambda _{23}\Gamma _{23}$.
We compute the biorthogonal spin Chern number for a wild range of $\lambda
_{23}$, as shown in Fig.~\ref{FigSCTR}(a). In the trivial insulator phase,
we have both $C_{\pm 1/2}=0$ as expected. Across the phase transition point,
the biorthogonal spin Chern number abruptly changes to $C_{\pm 1/2}=\pm 1$,
corresponding to the non-Hermitian QSH phase.

The other topological phase transition is ascribed to the non-Hermitian term
$i\lambda _{14}\Gamma _{14}$. Since we start from the QSH phase, we have $%
C_{\pm 1/2}=\pm 1$ when $|\lambda _{14}|$ is relatively small, as shown in
Fig.~\ref{FigSCTR}(b). In the trivial insulator phase, the biorthogonal spin
Chern numbers vanish $C_{\pm 1/2}=0$.

From these examples, we see that the biorthogonal spin Chern number
correctly characterizes the topological properties of the TR-symmetric
non-Hermitian Kane-Mele model. It provides an equivalent description as the
biorthogonal $\mathbb{Z}_{2}$ invariant. While a rigor proof of such
equivalence is hard to formulate, similar conclusion has been drawn in
Hermitian systems through the argument that the spin Chern numbers do not
contain more information than the $\mathbb{Z}_{2}$ invariant and vice versa
\cite{ProdanRobustness2009}. Similar arguments may be generalized to the
non-Hermitian cases.

\subsection{TR-broken QSH phase driven by non-Hermiticity}

In the main text, we use the biorthogonal spin Chern number to characterize
a topological phase transition from a trivial insulator phase to a TR-broken
QSH phase driven by the non-Hermitian term $i\lambda _{4}\Gamma ^{4}$, as
shown in Fig.~\ref{FigSC}.

\begin{figure}[h]
\centering
\includegraphics[width=0.48\textwidth]{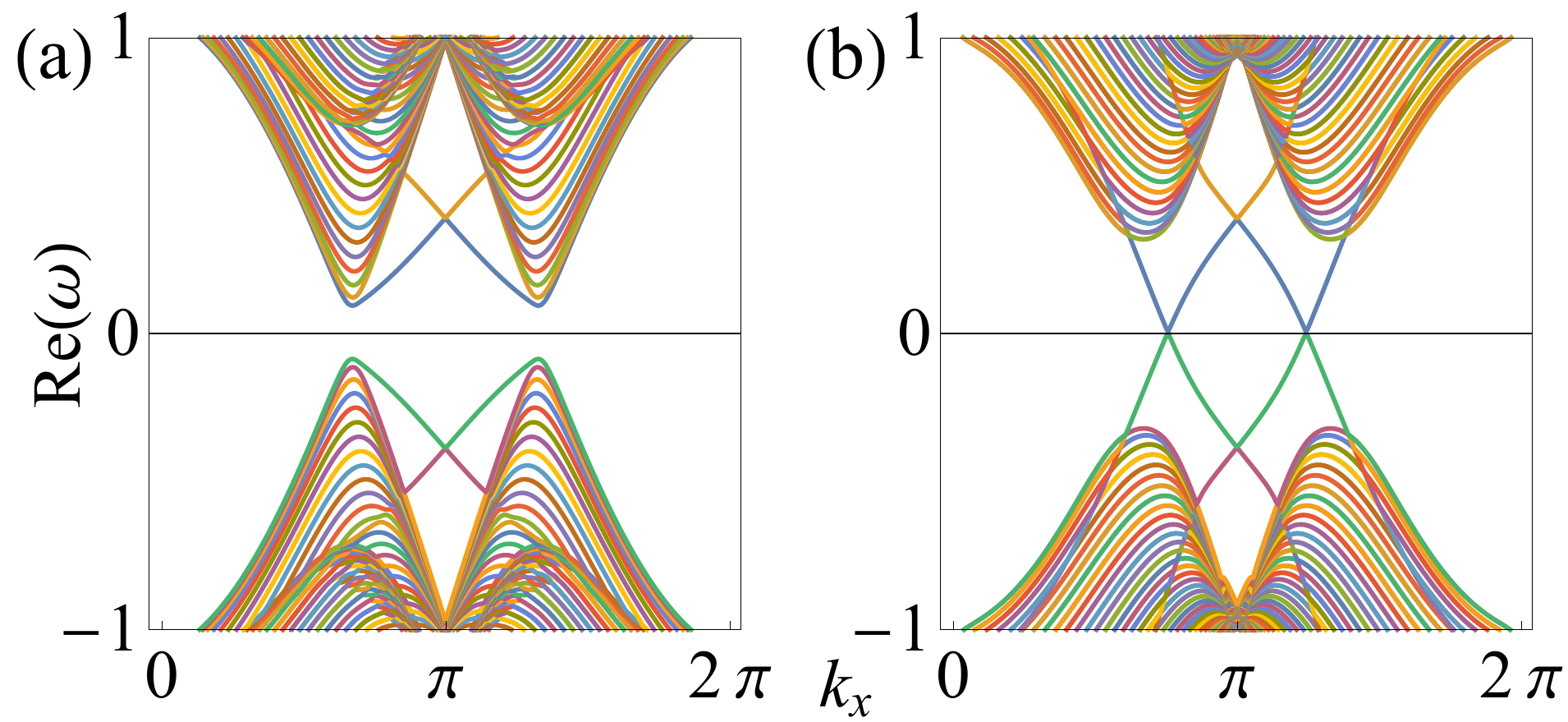}
\caption{TR-broken QSH phase driven by the non-Hermitian term $i\protect%
\lambda _{4}\Gamma ^{4}$. Open-boundary spectrum is plotted for (a) trivial
insulator phase ($\protect\lambda _{4}=0.1$) and (b) TR-broken QSH phase ($%
\protect\lambda _{4}=0.5$). Other parameters are the same as those in Figs.~%
\protect\ref{FigEdge}(a) and (b).}
\label{FigEdgeTRbk}
\end{figure}

In Fig.~\ref{FigEdgeTRbk}(a), we plot the open-boundary spectrum for $%
\lambda _{4}=0.1$, corresponding to a trivial insulator phase. The edge
states do not cross the band gap. With the increasing non-Hermitian
strength, the system enters the topological region. An example of $\lambda
_{4}=0.5$ is displayed in Fig.~\ref{FigEdgeTRbk}(b), where a typical edge
configuration for the QSH phase in the Kane-Mele model is found. The edge
states are consistent with the prediction from the biorthogonal spin Chern
number, demonstrating the bulk-edge correspondence.

\subsection{IQH phases in TR-broken non-Hermitian QSH insulators}

Previous study in Hermitian systems has incorporated the spin Chern number
to characterize the phase transition from a TR-broken QSH phase to an IQH
phase \cite{YangTime2011}. Here, we show that such a topological phase
transition still exists in non-Hermitian systems and their topological
properties are characterized by the biorthogonal spin Chern number.

\begin{figure}[h]
\centering
\includegraphics[width=0.48\textwidth]{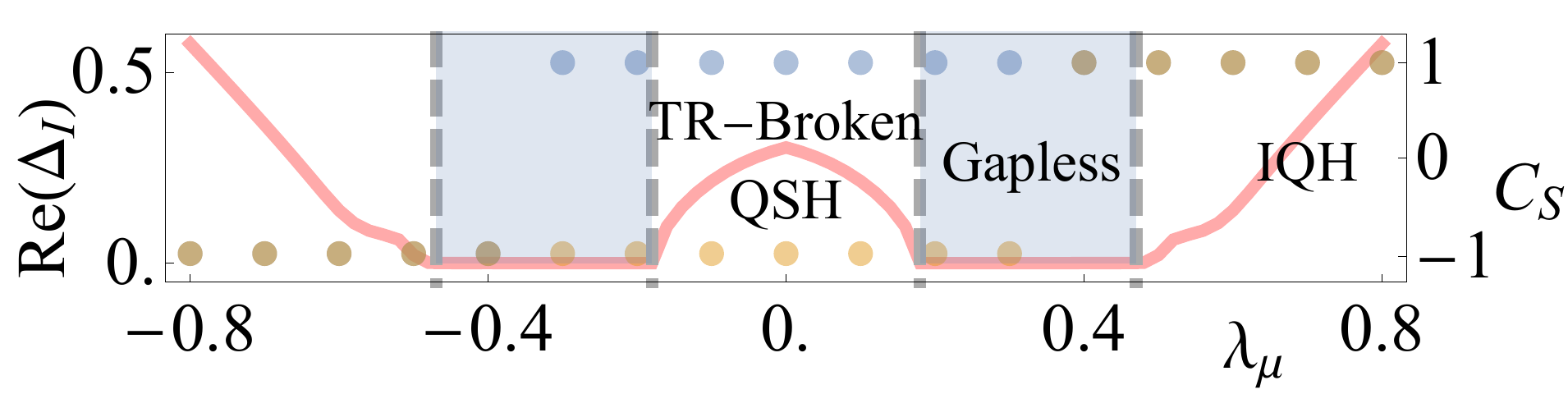}
\caption{Biorthogonal spin Chern number as the topological invariant for
TR-broken non-Hermitian QSH insulators. The topological phase transition is
induced by a real exchange field $\protect\lambda _{\protect\mu }\Gamma _{34}
$ and a small non-Hermitian term $i0.05\Gamma _{14}$ is included. Other
parameters are the same as those in Figs.~\protect\ref{FigEdge}(a,b).}
\label{FigSCSupp}
\end{figure}

We consider a non-Hermitian Kane-Mele model with a small non-Hermitian term $%
i\lambda _{14}\Gamma _{14}$ and a real exchange field $\lambda _{\mu}\Gamma
_{\mu}$. The biorthogonal spin Chern number with varying exchange field is
plotted in Fig.~\ref{FigSCSupp}. In the TR-broken QSH phase, the
biorthogonal spin Chern number remains non-trivial $C_{\pm 1/2}=\pm 1$. With
increasing exchange field, we observe a gapless phase and finally a
non-Hermitian IQH phase, where both spin sectors have the same biorthogonal
spin Chern number depending on the sign of the exchange field $C_{\pm 1/2}=%
\text{sign}(\lambda _{\mu})$. The spin Hall current vanishes but the charge
Hall conductance is quantized to $2$ (not strictly due to the Rashba
spin-orbit interaction and non-Hermitian effects).

\end{document}